# Evolution of Primordial Dark Matter Planets in the Early Universe

**Kiren O V**[1], **Kenath Arun***[1], **C Sivaram**[2]

[1]Department of Physics and Electronics, CHRIST (Deemed to be University), Bengaluru - 560029, India

[2]Indian Institute of Astrophysics, Bangalore - 560 034, India

**Abstract:** In a recent paper we had discussed possibility of DM at high redshifts forming primordial planets composed entirely of DM to be one of the reasons for not detecting DM (as the flux of ambient DM particles would be consequently reduced). In this paper we discuss the evolution of these DM objects as the Universe expands. As Universe expands there will be accretion of DM, helium and hydrogen layers (discussed in detail) on these objects. As they accumulate more and more mass, the layers get heated up leading to nuclear reactions which burn H and He when a critical thickness is reached. In the case of heavier masses of these DM objects, matter can be ejected explosively. It is found that the time scale of ejection is smaller than those from other compact objects like neutron stars (that lead to x-ray bursts). These flashes of energy could be a possible observational signature for these dense DM objects.

---

*Corresponding author:

e-mail: kenath.arun@christuniversity.in

## 1. Introduction

Dark matter is theorized as one of the basic constituents of the Universe, almost five times more abundant than ordinary matter. Many astronomical measurements have confirmed the existence of dark matter, leading to experiments worldwide like XENON1T experiment (Aprile et al., 2012; Undagoitia and Raunch, 2016) to directly observe dark matter particles. Till now the interaction of these particles with ordinary matter has proven to be so feeble that they have escaped direct detection (Arun et al., 2017). The DM effects can be, in principle and at least partially, be explained through the framework of extended gravity which has been considered earlier (Corda, 2009; 2018).

In the cosmic structure formation, the lightest objects would have formed first, i.e. the structure formation is a bottom-up scenario. It is of interest to note that the earliest objects to form could perhaps have been primordial planets dominantly composed of DM. It is not implied that all the DM particles go into forming such DM objects. The formation of such objects and their presence in large numbers in our galaxy could significantly reduce the number of free DM particles moving around in the Universe. The typical mass of such objects, made up mostly of DM particles of mass $m_D$, is given by (Sivaram and Arun 2011; Sivaram 1994)

$$M = \frac{M_{Pl}^3}{m_D^2} \qquad (1)$$

where the Planck mass is given by $M_{Pl} = \left(\hbar c/_G\right)^{1/2} \approx 2 \times 10^5 g$.

If we consider the mass of DM particles to be $60GeV$, favoured from the detection of excess of gamma rays from the galactic centre, attributed to the decay of $60GeV$ DM particles (Huang et al., 2016), the mass of the DM object works out to be $10^{29} g$ which is the mass of Neptune. This is the upper mass limit. The radius of the DM object is given by (Sivaram and Arun, 2011):

$$R = \frac{92\hbar^2}{G m_D^{8/3} M^{1/3}} \qquad (2)$$

where $m_D \sim 60GeV$ is the DM particle mass (Gelmini et al., 2006).

As the density of these objects fall of as $M^2$, the objects formed at later epochs would have a lower mass. If we consider the object density at a value 100 times the ambient density, say at $z = 10$, we get a lower mass limit of the object as $\sim 10^{14}$ g (typical asteroid mass). So the mass range of these DM objects will be from $10^{14}$ g (asteroid mass) extending to the mass of Neptune.

These objects could have formed in the early epochs of the Universe (when local DM density was much higher) and be in existence even now. Existence of (baryonic) primordial planets have been considered earlier by many authors (Shchekinov et al., 2013; Wickramasinghe et al., 2012). In the recent paper (Sivaram et al., 2019) we had discussed possibility of DM at high redshifts forming primordial planets composed entirely of DM to be one of the reasons for not detecting DM as the flux of ambient DM particles would be consequently reduced.

In Sivaram et al. (2016), we had proposed that the hypothesised Planet 9, in our solar system (Batygin and Brown, 2016; Trujillo and Sheppard, 2014), could indeed be such a DM planet, with a mass about that of Neptune. This might explain why it has not been visibly detected so far. Here we discuss the evolution of these DM planets formed in the early Universe, with the gravitational accretion of ambient hydrogen and helium by these objects.

## 2. Evolution of DM planets by mass accretion

The DM planets after their formation over a period of time would have accreted some mass. The rate at which mass is accreted for a body of mass M is given by the Bondi accretion rate.

$$\dot{M} = 4\pi R_{acc}^2 \rho V \quad (3)$$

where $\rho$ is the ambient density of medium, $R_{acc}$ is the radius of accretion given by, $R_{acc} = \left(\frac{2GM}{V^2}\right)^{\frac{1}{2}}$. The velocity $V$ of the accreted ambient particles is given by, $V^2 = c_s^2 + v^2$. Here $c_s$is the velocity of sound in the medium given by $c_s = \sqrt{\Upsilon R_g T}$, $\Upsilon$ is the ratio of specific heats, $R_g$ is the universal gas constant. Here $v$ is given by, $v = v_{amb} + v_{esc}$, where $v_{amb}$ is the velocity of the background ambient particles being accreted (with respect to the accreting object at rest). $v_{amb}$ is of the order of $100 km/s$.

and, $$v_{esc} = \left(\frac{2GM}{R}\right)^{\frac{1}{2}} \quad (4)$$

is the escape velocity from the DM object of radius $R$ (which would be the velocity at which the particles will be hitting the surface), $G$ is gravitational constant and $M$ is the mass of the object.

The accretion shock does not appear for the DM particles and hence $c_s$ is neglected and therefore in this case, $V = v$. The effects of dynamical friction (Chandrashekar, 1942) scales as $\rho/V^2$, and here $\rho$ being very small and $V$ being very high, it turns out that for the presently discussed situation this effect is quite negligible (Sivaram and Arun, 2014).

The DM particles are heavier compared to the ambient hydrogen and helium atoms, and are accreted much earlier since they are not coupled to the background radiation. The hydrogen and helium atoms are accreted after decoupling from the background radiation, with helium getting accreted before hydrogen since it decoupled earlier (ionization temperature of He being higher) (Switzer et al., 2008). During the accretion of hydrogen and helium, the shock waves change the densities of these gases as given by Rankine-Hugoniot equation (Zel'dovich, 1970).

$$\frac{\rho_2}{\rho_1} = \frac{\Upsilon+1}{\Upsilon-1} \quad (5)$$

where $\Upsilon$ is the ratio of specific heats and $\rho_1$ and $\rho_2$ are the densities of gas before and after the accretion shock.

The $c_s$ values of hydrogen and helium are calculated as $2.35 \times 10^6 cm/s$ and $2.78 \times 10^6 cm/s$ for a temperature of about $10^4 K$ (after recombination). This speed of sound is an order less than the ambient velocity and smaller compared to the escape velocity from the heavier masses. Hence accretion shock effects are negligible for heavier mass DM objects. After the recombination epoch, hydrogen was formed at redshifts of $z = 1000$. Since the ionization energy of helium is greater than that of hydrogen as it recombines earlier and this takes place around redshifts of $z = 3000$.

Hence these DM planets are expected to have a layer of DM particles, followed by successive layers of helium and hydrogen. $v_{esc}$ is important for higher mass DM objects and $v_{amb}$ is assumed to be same for all objects. From equation (3), the mass accreted per unit time for DM particles is:

$$dM = 4\pi R(t)^2 \rho_{DM}(t) v \, dt \quad (6)$$

As the Universe expands the ambient density changes with time, and the sizes of these planets changes with increase in their mass. The ambient densities of DM, H and He atoms varies with time as the Universe expands. From $z = 10$ (formation of earliest galaxies) to $z = 1$, the density of the DM as well as hydrogen and helium present within the galaxies become more dominant than the intergalactic ambient density. On the one hand we have the time-dependent (epoch dependent) background gas density given by (Rebecca et al., 2020; Sivaram et al., 2020):

$$\rho(t) = \rho_0(t_0) \left(\frac{t_0}{t}\right)^2 \quad (7)$$

where $\rho_0$ is the density at early epoch $t_0$.

On the other hand from $z = 1$ to present epoch, we have more or less constant density of these entities (i.e. DM, hydrogen and helium) within individual galaxies. The density of hydrogen and helium will now be 1000 orders less than its density at the earlier epochs. We assume that most of the planetary DM objects are now part of the galaxies (within the galaxies).

Using equations (2) and (7) in equation (6), the accreted mass over the complete epoch (from $t_0$ to present $t$) is given by:

$$\int_{M_0}^{M} M^{2/3}\, dM = 4\pi v t_0^{\,2} \left(\frac{92\hbar^2}{G m_D^{\,8/3}}\right)^2 \rho_0 \int_{t_0}^{t} \frac{1}{t^2}\, dt \tag{8}$$

where $M_0$ is the initial mass of the object. Equation (8) integrates to:

$$M^{5/3} = M_0^{\,5/3} + 4\pi v \rho_0 \left(\frac{92\hbar^2}{G m_D^{\,8/3}}\right)^2 t_0 \tag{9}$$

The accreted mass of H and He is given by a similar relation to equation (6) with $\rho$ now being the corresponding ambient density of these atoms.

The total mass accreted on the dark matter planet (whose mass varies from Neptune mass to asteroid mass) is tabulated for accretion of DM, He, and H separately in table 1(as calculated using equations (3) to (9)). ($\rho_0$ is the density at $t_0 = 10^{17} s$. At $Z = 3000$, $\rho_{0(DM)} = 5.4 \times 10^{-20} g/cc$, $\rho_{0(He)} = 2 \times 10^{-21} g/cc$. At $Z = 1000$, $\rho_{0(H)} = 3 \times 10^{-22} g/cc$)

Table 1: Velocity of the accreted particles and total mass of the ambient DM particles and H and He atoms accreted by the DM planets of various masses

| **Mass of Planet (g)** | **velocity (cm/s)** | **Mass of DM accumulated (g)** | **Mass of H accumulated (g)** | **Mass of He accumulated (g)** |
|---|---|---|---|---|
| $10^{29}$ | $1.08\text{x}10^{7}$ | $1\text{x}10^{29}$ | $7.65\text{x}10^{23}$ | $1.28\text{x}10^{24}$ |
| $10^{28}$ | $1.08\text{x}10^{7}$ | $1\text{x}10^{28}$ | $1.72\text{x}10^{21}$ | $2.89\text{x}10^{21}$ |
| $10^{27}$ | $1.08\text{x}10^{7}$ | $1\text{x}10^{27}$ | $4.31\text{x}10^{18}$ | $7.24\text{x}10^{18}$ |
| $10^{26}$ | $1.08\text{x}10^{7}$ | $1\text{x}10^{26}$ | $1.56\text{x}10^{16}$ | $2.62\text{x}10^{16}$ |
| $10^{25}$ | $1.08\text{x}10^{7}$ | $1\text{x}10^{25}$ | $9.64\text{x}10^{13}$ | $1.62\text{x}10^{14}$ |
| $10^{24}$ | $1.08\text{x}10^{7}$ | $1\text{x}10^{24}$ | $8.35\text{x}10^{11}$ | $1.40\text{x}10^{12}$ |
| $10^{23}$ | $1.08\text{x}10^{7}$ | $1\text{x}10^{23}$ | $8.11\text{x}10^{9}$ | $1.36\text{x}10^{10}$ |
| $10^{22}$ | $1\text{x}10^{7}$ | $1\text{x}10^{22}$ | $8.03\text{x}10^{7}$ | $1.35\text{x}10^{8}$ |
| $10^{21}$ | $1\text{x}10^{7}$ | $1\text{x}10^{21}$ | $8.03\text{x}10^{5}$ | $1.35\text{x}10^{6}$ |
| $10^{20}$ | 1x$10^{7}$ | 1x$10^{20}$ | 8.03x$10^{3}$ | 1.35x$10^{4}$ |
| $10^{19}$ | 1x$10^{7}$ | 1x$10^{19}$ | 8.03x$10^{1}$ | 1.35x$10^{2}$ |
| $10^{18}$ | 1x$10^{7}$ | 1x$10^{18}$ | 8.03x$10^{-1}$ | 1.35x$10^{0}$ |
| $10^{17}$ | 1x$10^{7}$ | 1x$10^{17}$ | 8.03x$10^{-3}$ | 1.35x$10^{-2}$ |
| $10^{16}$ | 1x$10^{7}$ | 1.01x$10^{16}$ | 8.03x$10^{-5}$ | 1.35x$10^{-4}$ |
| $10^{15}$ | 1x$10^{7}$ | 2x$10^{15}$ | 8.03x$10^{-7}$ | 1.35x$10^{-6}$ |
| $10^{14}$ | 1x$10^{7}$ | 2x$10^{14}$ | 8.03x$10^{-9}$ | 1.35x$10^{-8}$ |

## 3. Formation of accreted layers on the DM planets and their dynamics

As more and more mass gets accreted, the temperatures of accreted hydrogen and helium layers start increasing, whereas accreted DM will not be heated up. The temperatures of the accreted layers of hydrogen and helium on the DM planet is given by:

$$T = \frac{GMm}{Rk_B} \tag{10}$$

where $M$ is mass of planet, $m$ is mass of H or He atom, $R$ is radius of planet and $k_B$ is Boltzmann constant.

We have tabulated the temperatures (table 2) of these hydrogen and helium layers on the DM planet and found that the temperature is high enough for heavier DM planets to have nuclear reactions. But these nuclear reactions can happen only if these high temperatures are retained for sufficient time for the reactions to occur. The time scale of cooling is given as:

$$t = \frac{M_{acc} R_g T}{\sigma T^4 A} \tag{11}$$

where $M_{acc}$ is mass accreted, $R_g$ is gas constant, $T$ is temperature of gas layer, $\sigma$ is Stefan Boltzmann constant and $A$ is the surface area of accreting planet, ($A = 4\pi R^2$, $R$ as given in equation(2)).

Table 2: Temperature and the time scale of cooling for the H and He layers

| **M (g)** | $T_H$ **(K)** | $T_{He}$ **(K)** | $t_H$ **(s)** | $t_{He}$ **(s)** |
|---|---|---|---|---|
| $10^{29}$ | 5.38 x $10^{9}$ | 2.14 x $10^{10}$ | 2.55 x $10^{-3}$ | 6.68 x $10^{-5}$ |
| $10^{28}$ | 2,52 x $10^{8}$ | 1.01 x $10^{9}$ | 1.22 x $10^{-2}$ | 3.20 x $10^{-4}$ |
| $10^{27}$ | 1.15 x $10^{7}$ | 4.60 x $10^{7}$ | 6.70 x $10^{-2}$ | 1.76 x $10^{-3}$ |
| $10^{26}$ | 1.20 x $10^{6}$ | 4.80 x $10^{6}$ | 4.68 x $10^{-2}$ | 1.23 x $10^{-3}$ |
| $10^{25}$ | 1.20 x $10^{6}$ | 4.80 x $10^{6}$ | 6.36 x $10^{-5}$ | 1.67 x $10^{-6}$ |
| $10^{24}$ | 1.20 x $10^{6}$ | 4.80 x $10^{6}$ | 1.15 x $10^{-7}$ | 3.02 x $10^{-9}$ |
| $10^{23}$ | 1.20 x $10^{6}$ | 4.80 x $10^{6}$ | 2.44 x $10^{-10}$ | 6.39 x $10^{-12}$ |
| $10^{22}$ | 1.20 x $10^{6}$ | 4.80 x $10^{6}$ | 5.30 x $10^{-13}$ | 1.39 x $10^{-14}$ |
| $10^{21}$ | 1.20 x $10^{6}$ | 4.80 x $10^{6}$ | 1.11 x $10^{-15}$ | 2.90 x $10^{-17}$ |
| $10^{20}$ | 1.20 x $10^{6}$ | 4.80 x $10^{6}$ | 2.41 x $10^{-18}$ | 6.33 x $10^{-20}$ |
| $10^{19}$ | 1.20 x $10^{6}$ | 4.80 x $10^{6}$ | 5.30 x $10^{-21}$ | 1.39 x $10^{-22}$ |
| $10^{18}$ | 1.20 x $10^{6}$ | 4.80 x $10^{6}$ | 1.11 x $10^{-23}$ | 2.90 x $10^{-25}$ |
| $10^{17}$ | 1.20 x $10^{6}$ | 4.80 x $10^{6}$ | 2.41 x $10^{-26}$ | 6.33 x $10^{-28}$ |
| $10^{16}$ | 1.20 x $10^{6}$ | 4.80 x $10^{6}$ | 5.30 x $10^{-29}$ | 1.39 x $10^{-30}$ |
| $10^{15}$ | 1.20 x $10^{6}$ | 4.80 x $10^{6}$ | 1.11 x $10^{-31}$ | 2.90 x $10^{-33}$ |
| $10^{14}$ | 1.20 x $10^{6}$ | 4.80 x $10^{6}$ | 2.41 x $10^{-34}$ | 6.33 x $10^{-36}$ |

**4. Nuclear reactions and ejection of mass from H and He layers**

Helium burning requires a temperature of 200 million kelvin ($2 \times 10^{8} K$), and H burning requires few million degrees ($\sim 10 - 30 \times 10^{6} K$) (Burbidge et al., 1957; Caughlan and Fowler., 1988). So only the objects having masses $10^{29}\ g$ and $10^{28}\ g$ can fuse He, objects with $10^{27}\ g$ can fuse H (to He) and objects with masses $10^{26}\ g$ and below cannot undergo nuclear reactions. However for masses from $10^{25}\ g$ to $10^{22}\ g$, since the temperatures are not high enough and moreover the cooling times are smaller, the conditions are not sufficient for nuclear reactions to happen. These objects emit energy from the heat accumulated by the accreting layers. The energy radiated by these lower masses is tabulated in Table 6.

For lower masses ($10^{21}\ g$ and below), the temperature of these gases cools down very fast leaving no time for nuclear reaction to occur. But however for heavier masses there are reactions possible where H and He could get converted into heavier elements. (Sivaram et al., 2014)

For the reactions to happen a certain thickness of hydrogen and helium layers should be accumulated above the DM planets' surfaces. The potential energy of the accreted mass is given by:

$$\Delta U = \frac{GMM_{acc}}{R} \qquad (12)$$

where $M_{acc}$ is the mass accreted on the planet. This potential energy keeps on increasing with the mass accumulation leading to an increase in the pressure energy (thermal energy density) given by:

$$P = \rho R_g T \tag{13}$$

where $R_g$ is the gas constant, $T$ is the temperature of the layer. This thermal energy density must be equal to the potential energy density, (i.e. $\rho R_g T = \rho g h$) and thus we obtain the thickness of the layer accreted on the planet as,

$$h = \frac{R_g T}{g} \tag{14}$$

Here $g$ is acceleration due to gravity of the planet given by, $g = GM/R^2$. To estimate required thickness of layer i.e. $h$, $T$ is taken as $200 \times 10^6 K$ for He and $30 \times 10^6 K$ for H to undergo fusion. By using equations (12) and (13) and differentiating with time we get:

$$\dot{\rho} = \frac{GM\dot{M}_{acc}}{4\pi R^3 h R_g T} \tag{15}$$

where $\dot{\rho}$ is the rate of change of density of the mass accumulated and $\dot{M}_{acc}$ is rate of mass accreted. The density of the accreted layers keeps increasing over time which increases the gravitational pressure. The gravitational pressure is given by:

$$P = \frac{GM_{DM}M_{acc}}{R^4} \tag{16}$$

The radiation pressure exerted by the heated up accreted gas also increases and when gravitational pressure and corresponding radiation pressure reaches a maximum, the layer is ejected out into space. The maximum value of gravitational pressure is obtained by equating it with the radiation pressure, i.e.:

$$\frac{GM_{DM}M_{acc}}{R^4} \leq \frac{\sigma T^4}{c} \tag{17}$$

where σ is Stefan-Boltzmann constant, c is the speed of light.

Equation (17) implies that if $M_{acc}$ exceeds some value, radiation pressure will dominate and could lead to ejection of mass. The mass accreted for a thickness of layer $h$ is given by:

$$M_{acc} = 4\pi R^2 h \rho \tag{18}$$

By using equations (17) and (18) we can obtain the maximum density of the accreted layers as given by:

$$\rho = \frac{\sigma T^4 R^2}{GM_{DM} 4\pi h c} \tag{19}$$

Thus the mass of the accreted hydrogen and helium layers for the estimated thickness on the heavier DM objects is tabulated (using the above equations) in table 3. The thickness of layer accumulated for H and He for different of mass of the object is plotted in figure 1.

Table 3: Height (thickness), density, and mass of the accreted layers on the heavier DM objects

| $\boldsymbol{M_{DM}(g)}$ | **Helium layer** | | | **Hydrogen layer** | | |
|---|---|---|---|---|---|---|
| | $\boldsymbol{h\ (cm)}$ | $\boldsymbol{\rho\ (g/cc)}$ | $\boldsymbol{M_{acc}\ (g)}$ | $\boldsymbol{h\ (cm)}$ | $\boldsymbol{\rho\ (g/cc)}$ | $\boldsymbol{M_{acc}\ (g)}$ |
| $10^{29}$ | $4.5\text{x}10^{2}$ | $2.37\text{x}10^{9}$ | $3.01\text{x}10^{21}$ | $3.15\text{x}10^{1}$ | $1.35\text{x}10^{8}$ | $1.2\text{x}10^{19}$ |
| $10^{28}$ | $2.05\text{x}10^{4}$ | $1.17\text{x}10^{4}$ | $3.09\text{x}10^{18}$ | $1.43\text{x}10^{3}$ | $6.50\text{x}10^{2}$ | $1.2\text{x}10^{16}$ |
| $10^{27}$ | $9.8\text{x}10^{5}$ | $5.05\text{x}10^{-2}$ | $3.05\text{x}10^{15}$ | $6.86\text{x}10^{4}$ | $2.82\text{x}10^{-3}$ | $1.19\text{x}10^{13}$ |
| $10^{26}$ | $4.5\text{x}10^{7}$ | $5.99\text{x}10^{-6}$ | $7.61\text{x}10^{13}$ | $3.15\text{x}10^{6}$ | $3.34\text{x}10^{-7}$ | $2.97\text{x}10^{11}$ |

The Eddington luminosity of a body of mass $M$ is given by:

$$L = \frac{4\pi G M m_p c}{\sigma_T} \quad (20)$$

where $\sigma_T$ is Thompson scattering cross section for electron and $m_p$ is the mass of proton.

For a mass of $10^{29}\ g$, the Eddington luminosity is $10^{34}\ erg/s$. The energy released when $1\ g$ of H fuses to He is $3 \times 10^{18}\ erg$. So for a luminosity of $10^{34}\ erg/s$, the corresponding mass required is $10^{16}\ g$. So any accumulated mass more than $10^{16}\ g$ will be ejected out with Eddington luminosity. Similarly the energy released in the nuclear fusion of He is $10^{18}\ erg/s$ and hence mass required for Eddington luminosity is again $10^{16}\ g$. We have tabulated the Eddington luminosities and the critical mass limit for ejection of these layers of H and He in table 4.

For larger masses the mass accreted is more than the limiting mass rate. So these larger masses will eject this excess mass with Eddington luminosity in the timescale given by:

$$t_{eject} = \frac{Mass\ accreted}{limiting\ mass\ rate} \quad (21)$$

The lower mass DM objects which accrete masses lower than the critical mass required for fusion will continue to release energy with luminosity given by:

$$L = 4\pi R^2 \sigma T^4 \quad (22)$$

where $R$ is radius of planet and $T$ is the temperature of layer. The total energy released by the mass accreted is given by:

$$\text{Energy released} = \text{Mass accreted} \times \text{Energy released per gram} \quad (23)$$

The luminosity of the ejected mass is given by:

$$\text{Luminosity} = \frac{\text{Energy released}}{\text{Time scale of ejection}} \quad (24)$$

These are tabulated in table 5.

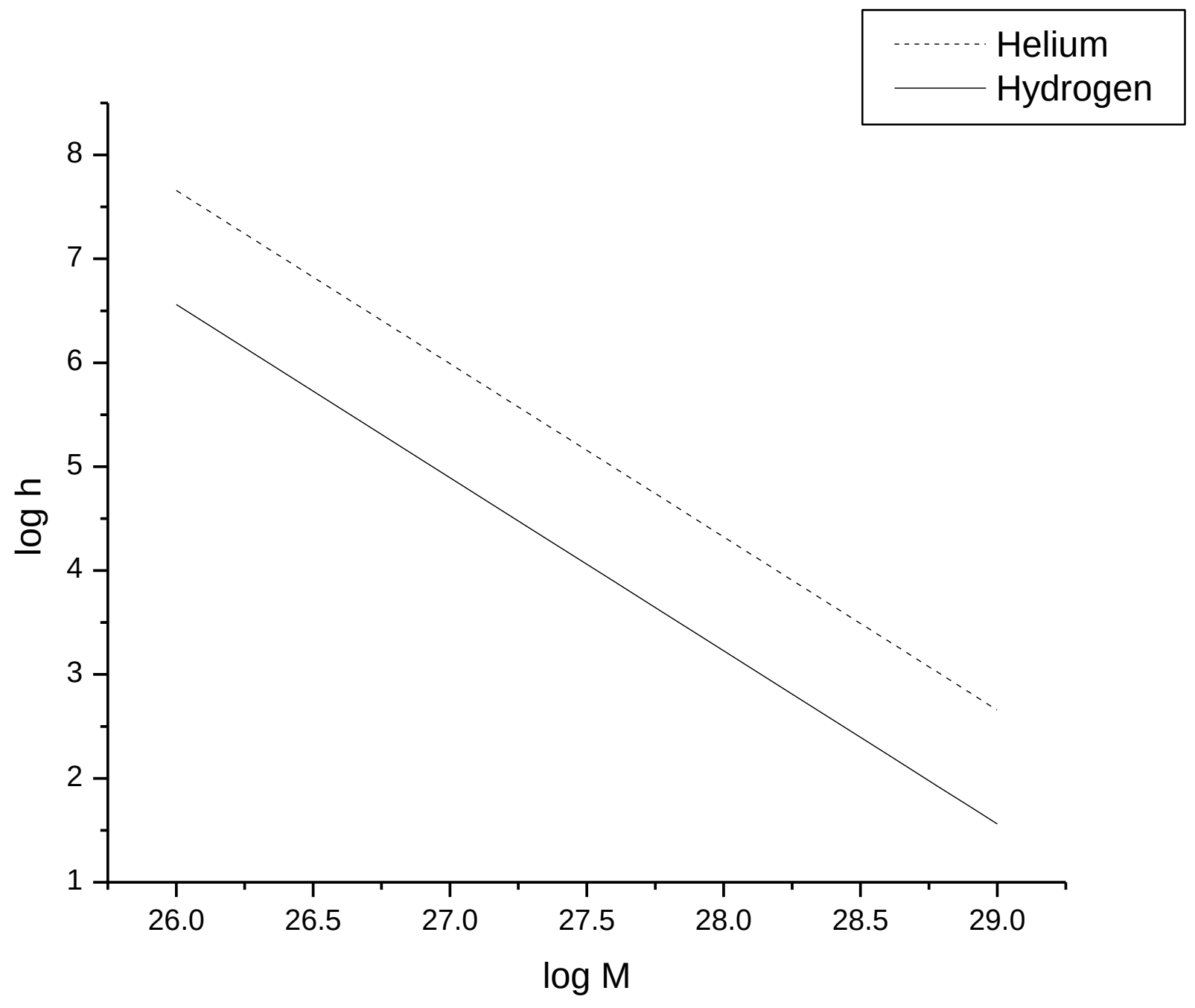

Figure 1: Thickness of layer accumulated for H and He for different of mass of the object

Table 4: Critical mass of the layer (required to trigger nuclear reaction) and Eddington luminosities of H and He layers accreted on DM objects of heavier mass

| $M_{DM}(g)$ | $L_{Edd}(erg/s)$ | Helium layer | | Hydrogen layer | |
|---|---|---|---|---|---|
| | | **Limiting mass rate (g/s)** | **Mass accreted (g)** | **Limiting mass rate (g/s)** | **Mass accreted (g)** |
| $10^{29}$ | $10^{34}$ | $10^{16}$ | $3.01 \times 10^{21}$ | $10^{16}$ | $1.2 \times 10^{19}$ |
| $10^{28}$ | $10^{33}$ | $10^{15}$ | $3.09 \times 10^{18}$ | $10^{15}$ | $1.2 \times 10^{16}$ |
| $10^{27}$ | $10^{32}$ | $10^{14}$ | $3.05 \times 10^{15}$ | $10^{14}$ | $1.19 \times 10^{13}$ |
| $10^{26}$ | $10^{31}$ | $10^{13}$ | $7.61 \times 10^{13}$ | $10^{13}$ | $2.97 \times 10^{11}$ |

Table 5: Energy released by the layers and the time scale of these ejections

| $M_{DM}(g)$ | Helium layer | | | Hydrogen layer | | |
|---|---|---|---|---|---|---|
| | **Energy released (erg)** | **Time of ejection (s)** | **Luminosity (ergs/s)** | **Energy released (erg)** | **Time of ejection (s)** | **Luminosity (ergs/s)** |
| $10^{29}$ | $3.01 \times 10^{39}$ | $3.01 \times 10^{5}$ | $10^{34}$ | $3.6 \times 10^{37}$ | $1.2 \times 10^{3}$ | $3 \times 10^{34}$ |
| $10^{28}$ | $3.09 \times 10^{36}$ | $3.09 \times 10^{3}$ | $10^{33}$ | $3.6 \times 10^{34}$ | $1.2 \times 10^{1}$ | $3 \times 10^{33}$ |
| $10^{27}$ | $3.05 \times 10^{33}$ | $3.05 \times 10^{1}$ | $10^{32}$ | $3.57 \times 10^{31}$ | $1.19 \times 10^{-1}$ | $3 \times 10^{32}$ |
| $10^{26}$ | $7.61 \times 10^{31}$ | 7.61 | $10^{31}$ | $8.91 \times 10^{29}$ | $2.97 \times 10^{-2}$ | $3 \times 10^{31}$ |

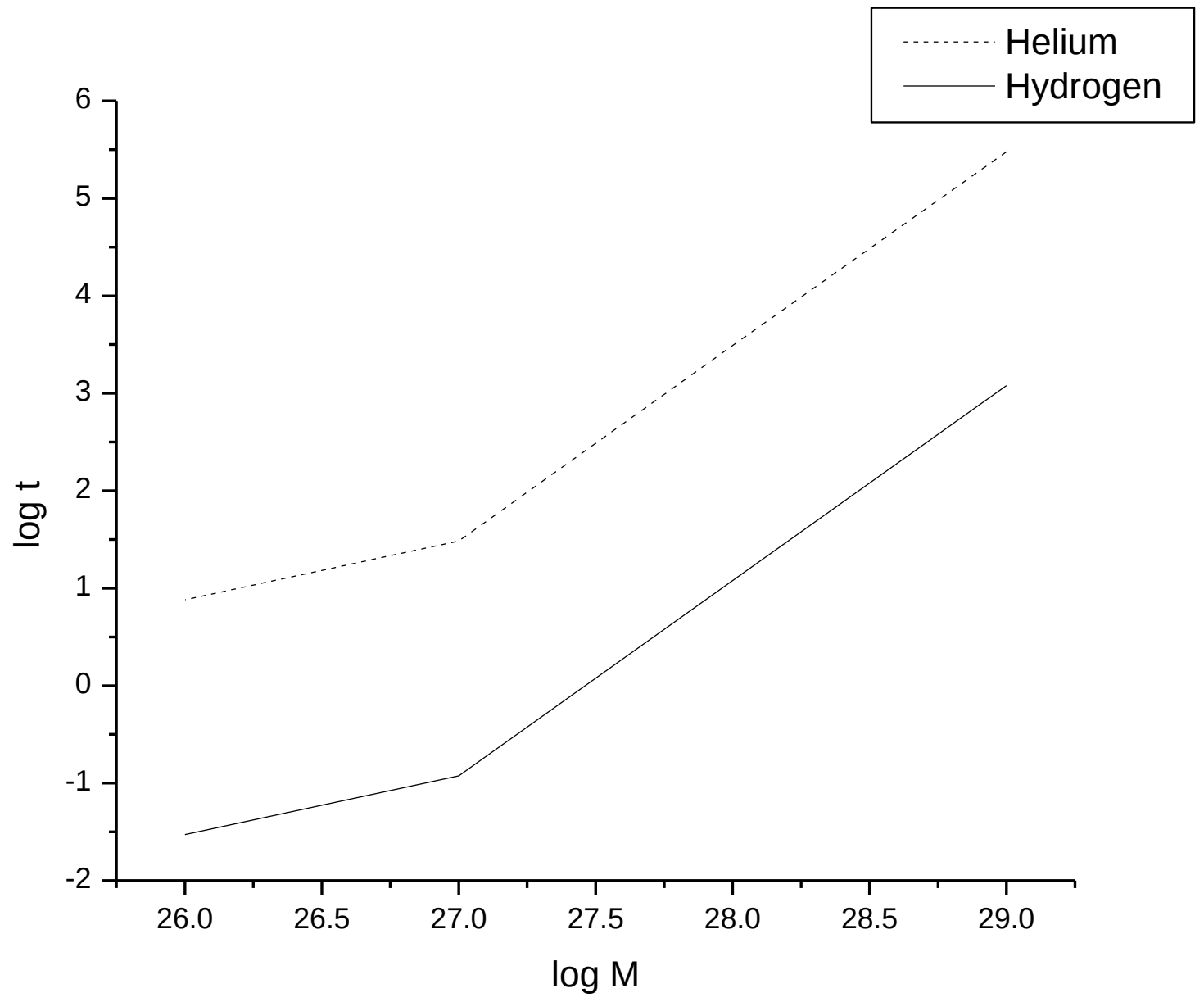

Figure 2: The time scale of ejection of the accumulated layers of H and He for different mass of the DM object

The Luminosity ($L$) and Radius ($R$) of the objects is calculated by equation (22) and equation (2), respectively. Here $T$ is the temperature of layer, $t$ is the time scale of cooling (the values are from table 2). The Energy radiated is given by $E =$ Luminosity $\times$ time of cooling. The time scale of ejection of the accumulated layers of H and He for different mass of the DM object is plotted in figure 2.

Table 6: Thermal energy released by H and He layers accumulating on lower mass objects

| $M(g)$ | $R(cm)$ | Hydrogen layer | | | | Helium layer | | | |
|---|---|---|---|---|---|---|---|---|---|
| | | $T_H\ (K)$ | $L(erg/s)$ | $t_H\ (s)$ | $E\ (erg)$ | $T_{He}\ (K)$ | $L(erg/s)$ | $t_{He}\ (s)$ | $E\ (erg)$ |
| $10^{25}$ | $3.2\times10^{5}$ | $1.2\times10^{6}$ | $1.5\times10^{32}$ | $1.6\times10^{7}$ | $2.4\times10^{39}$ | $4.8\times10^{6}$ | $3.9\times10^{34}$ | $2.5\times10^{5}$ | $9.6\times10^{39}$ |
| $10^{24}$ | $7\times10^{5}$ | $1.2\times10^{6}$ | $7.2\times10^{32}$ | $3.3\times10^{5}$ | $2.4\times10^{38}$ | $4.8\times10^{6}$ | $1.8\times10^{35}$ | $5.2\times10^{3}$ | $9.6\times10^{38}$ |
| $10^{23}$ | $1.5\times10^{6}$ | $1.2\times10^{6}$ | $3.3\times10^{33}$ | $7.2\times10^{3}$ | $2.4\times10^{37}$ | $4.8\times10^{6}$ | $8.5\times10^{35}$ | $1.1\times10^{2}$ | $9.6\times10^{37}$ |
| $10^{22}$ | $3.2\times10^{6}$ | $1.2\times10^{6}$ | $1.5\times10^{34}$ | $1.6\times10^{2}$ | $2.4\times10^{36}$ | $4.8\times10^{6}$ | $3.9\times10^{36}$ | 2.48 | $9.6\times10^{36}$ |

## 5. Possible effects on CMB

The cosmic microwave background (CMB) will be distorted only if the energies of the nuclear reactions are released, around $z = 1000$ or earlier. However here, the reactions occur after the H and He layers are accreted on the DM object to a sufficient thickness. They start accreting these atoms only after recombination i.e. later than $z = 1000$. Again sufficient time elapses before the layers reach this thickness. So when the reactions and ejections do take place, it is well after the recombination era, implying that the CMB will not be affected. Moreover only the heavier mass objects in the range of $10^{26} - 10^{29} g$ undergo these reactions and the total energy released is only a very small fraction (several orders less) of the total energy associated with the CMB.

## 6. Conclusions

Following our earlier works on primordial planets composed of DM, we have considered their evolution as the Universe expands. This involves accretion of ambient DM, hydrogen and helium on these objects forming successive layers (after H and He recombine in the early Universe). The H and He layers get heated up and for the heavier DM objects, the layers after reaching a critical thickness could undergo nuclear reactions burning He and H. The luminosities are estimated. The time scales of ejection in case of Eddington luminosity reached by the layers is estimated. The ejections leading to flashes would be much less energetic than the bursts (X-ray bursts) from neutron stars, the corresponding time scales also being shorter. The above estimates could be typical signatures for future observations. A more detailed study is under way.